\documentclass[twocolumn, aps, prl, superscriptaddress,floatfix]{revtex4-1}

\usepackage{multirow}
\usepackage{epsfig}
\usepackage{amsmath}

\usepackage{latexsym}
\usepackage{feynmp}

\usepackage[normalem]{ulem}  
\usepackage[dvips]{color} 

\renewcommand\sout{\bgroup \color{red} \ULdepth=-.5ex \ULset}

\begin{document}

\title{Tribaryon configurations and the inevitable three nucleon repulsions at short distance}

\author{Aaron Park}
\email{aaron.park@yonsei.ac.kr}\affiliation{Department of Physics and Institute of Physics and Applied Physics, Yonsei
University, Seoul 03722, Korea}
\author{Woosung Park}
\email{diracdelta@yonsei.ac.kr}\affiliation{Department of Physics and Institute of Physics and Applied Physics, Yonsei
University, Seoul 03722, Korea}
\author{Su Houng Lee}
\email{suhoung@yonsei.ac.kr}\affiliation{Department of Physics and Institute of Physics and Applied Physics, Yonsei
University, Seoul 03722, Korea}

\begin{abstract}
 We decompose the tribaryon configuration in terms of SU(3) flavor and spin state and analyse their color-spin-flavor wave function following Pauli principle.  By comparing the color-color and color-spin interactions of compact tribaryon configuration  against the lowest three nucleon threshold within a constituent quark model,  we show that the three nucleon  configurations have to be repulsive at short distance for all possible quantum numbers and all values of the SU(3) symmetry breaking parameter.  Our work identifies the origin of the repulsive nuclear three body forces including the hyperons at short distance that are called for from phenomenological considerations starting from nuclear matter to the maximum mass of a neutron star.

\end{abstract}

\maketitle


The importance of three nucleon force  was discussed from the early days of nuclear physics\cite{Primakoff:1939zz,Fujita:1957zz} to address nuclear configurations beyond the deuteron\cite{Pieper:2001ap,Epelbaum:2002vt,Hammer:2012id}.
 Phenomenologically, the long distance part of the three nuclear force is obtained using the pion mediated interactions\cite{Coon:1978gr,Carlson:1983kq} which gives  the additional attraction necessary to fit the experimental $^3H$ binding energy.
 However, as one tries to go beyond the few body system and understand the saturation properties of nuclear matter\cite{Carlson:1983kq,Kievsky:2008es} and the compressibility consistent with experiments\cite{Sakuragi:2016jni},
 it becomes clear that  the three body nuclear force including those involving hyperons\cite{Nishizaki:2002ih} are repulsive at short distance.

Recently, there is a renewed interest in the three body nuclear force as there are compelling    evidences that  the  force should be  repulsive at short distance from analyzing the extrapolated dense nuclear matter equation of state (EOS) in the  neutron stars\cite{Lattimer:2006xb,Kojo:2017gxc}.  The
naive extrapolation of two body interaction fitted to hypernuclear data leads to $\Lambda$ hyperon condensation at 2 $\sim$ 3 times nuclear matter density in neutron stars, which leads to maximum neutron star mass of less than 1.5 M$_{\odot}$\cite{Bombaci:2016xzl}.  On the other hand the
recently observed neutron stars J1614-2230\cite{Demorest:2010bx}  and J0348-0432\cite{Antoniadis:2013pzd} have observed masses of $(1.97 \pm 0.04) {\rm M_{\odot}}$ and $(2.01 \pm 0.04) {\rm M_{\odot}}$ respectively.
A natural way to solve the "Hyperon Puzzle" is to introduce a repulsive short distance three body force including the hyperon which becomes more important at high density and naturally leads to a stiffer equation of state and forces the hyperons to appear at a much higher density\cite{Yamamoto:2014jga}.

In this work, we will show why the short distance part of the three nucleon interaction, including the hyperons, should be repulsive.  Our calculation is based on Pauli principle and the color-color and color-spin interaction for a decomposed color-flavor-spin wave function with broken flavor SU(3).   Specifically, we compare the stability of the compact tribaryon configuration
against three nucleon configuration of the lowest threshold within a constituent quark picture for all possible quantum numbers.
We find that for every quantum number, the tribaryon configuration is highly repulsive compared to the lowest three nucleon threshold even in the SU(3) flavor symmetry broken case.  In particular, we find repulsion even in the least repulsive configuration with one strange quark, suggesting that indeed the naive linear density extrapolated attraction leading to hyperon condensation will break down due to the three body repulsion between a hyperon and two nucleons.

Although the QCD origin for the short range repulsion for two nucleon systems were discussed in the quark-cluster model\cite{Oka:1980ax,Oka:1981ri} and recently from direct lattice calculations\cite{Ishii:2006ec,Aoki:2012tk}, the case for the three nucleon case has never been discussed before.   Our calculation is based on the basic  properties of the color-spin-flavor wave function using all possible two and three body quark interaction based on color and spin between s-wave quarks within the constituent quark model  and can also be used to estimate the phenomenological parameters for short distance repulsion.

{\it Interaction:}
In general, the s-wave quark interaction can be of color-color or color-spin interaction type.  Here, we do not consider the possible exchange of pseudoscalar bosons between quarks\cite{Glozman:1995fu}.
The specific form in the constituent quark model will involve spatial functions that include the confining and coulomb potential \cite{Bhaduri:1981pn,Park:2016mez}.
In this work, we will not work out a specific constituent quark model, but focus the discussions to compact configurations where all the s-wave quarks occupy the  size  of a baryon so that the spatial wave function will be the same for all quarks.    In such an approximation,
the matrix element for the interaction terms will be factorized into a spatial part and the color-spin-flavor dependent part.
 The spatial part of the interaction becomes universal for all quarks
so that the interaction part of the total Hamiltonian  for any physical state composed of $N$ quarks can be approximated as
\begin{eqnarray}
H_{color-color}& = & -C_{CC} \sum_{i<j} \lambda^c_i \lambda^c_j , \label{cc} \\
H_{color-spin}
&=&-C_{CS}\sum_{i<j} \frac{1}{m_i m_j} \lambda^c_i \lambda^c_j
{\sigma}_i\cdot{\sigma}_j  . \label{cs}
\end{eqnarray}
Here,   $m_i$ is the constituent quark mass of the $i$ quark. $C$'s are the constant that depend on the spatial part of the wave function, which we will take to be universal for all physical states composed of s-wave quarks.
The simplified model well reproduces the hyperfine splitting between bayrons\cite{Lee:2007tn} so that the stability between various quantum numbers can be analysed  from the color-color and color-spin matrix elements.

Let us first consider the color-color interaction of a tribaryon configuration for a given quantum number. Since both the multiquark configuration and the baryons are color singlet states, by calculating $( \sum_i^N \lambda_i^c )^2$, one obtains the following formula.
\begin{eqnarray}
\sum_{i<j} \lambda_i^a \lambda_j^a  & = & -\frac{1}{2}N \lambda_q^2,
\end{eqnarray}
where $\lambda_q^2=\frac{16}{3}$ for all quarks.  Hence, the color-color interaction of a compact multiquark configuration will just be the sum of individual baryons and will thus give negligible contribution to the static interaction  energy between baryons at short distance.

Next,  we investigate the color-spin interaction of the tribaryon given in Eq.~(\ref{cs}).
%
In the SU(3) flavor symmetric case, there is a well-known formula to calculate the color-spin interaction for a totally antisymmetric multiquark configuration\cite{Aerts:1977rw}.
\begin{align}
  K&=-\sum_{i<j} \lambda_i^c \lambda_j^c \sigma_i \cdot \sigma_j \nonumber \\
  &= N(N-10)+\frac{4}{3}S(S+1)+4C_F+2C_C,
\label{color-spin-formula}
\end{align}
where $N$ is the total number of quarks and $C_F=\frac{1}{4}\lambda^F \lambda^F$, which is the first kind of Casimir operator of flavor SU(3).
As is evident from this formula, the color-spin interaction has a non-linear dependence in $N$ and can contribute to either repulsion or attraction between baryons.

We define the static binding  energy of the tribaryon against the lowest three baryon threshold
 as follows:
\begin{align}
  E_B=H_{\mathrm{tribaryon}}-H_{\mathrm{baryon1}}-H_{\mathrm{baryon2}}-H_{\mathrm{baryon3}}. \label{binding}
\end{align}
As the spatial part of the wave function is universal, one notes that the $K$ factor defined in Eq.~(\ref{color-spin-formula}) is important
in the short distance part of the tribaryon interaction.

{\it Flavor state of the tribaryon:}
We first analyze the possible compact configurations of 9-quarks allowed within the Pauli principle.
The flavor-color-spin wave function of the tribaryon should be antisymmetric as the spatial configurations are taken to be symmetric. Since the color of the tribaryon should be in the   singlet state $[333]_C$, antisymmetry requires the remaining flavor-spin wave function to be in the conjugate representation $[333]_{FS}$.  Then, we can decompose the flavor-spin state in terms of SU(3) flavor and SU(2) spin symmetry.\\

$[333]_{FS}=[63]_F \otimes [63]_S + [54]_F \otimes [54]_S + [621]_F \otimes [54]_S + [531]_F \otimes [72]_S + [531]_F \otimes [63]_S + [531]_F \otimes [54]_S + [522]_F \otimes [63]_S + [441]_F \otimes [63]_S + [432]_F \otimes [81]_S + [432]_F \otimes [72]_S + [432]_F \otimes [63]_S + [432]_F \otimes [54]_S + [333]_F \otimes [9]_S + [333]_F \otimes [72]_S + [333]_F \otimes [63]_S$.\\

There are eight possible flavor states for tribaryon as follows.\\

\begin{small}
$\begin{tabular}{|c|c|c|c}
  \cline{1-3}
  \quad \quad & \quad \quad & \quad \quad  \\
  \cline{1-3}
  \quad \quad & \quad \quad & \quad \quad \\
  \cline{1-3}
  \quad \quad & \quad \quad & \quad \quad \\
  \cline{1-3}
  \multicolumn{4}{c}{$\mathbf{1}(S=\frac{3}{2},\frac{5}{2},\frac{9}{2})$}
\end{tabular}$,
$\begin{tabular}{|c|c|c|c|c}
  \cline{1-4}
  \quad \quad & \quad \quad & \quad \quad & \quad \quad & \quad \quad  \\
  \cline{1-4}
  \quad \quad & \quad \quad & \quad \quad \\
  \cline{1-3}
  \quad \quad & \quad \quad \\
  \cline{1-2}
  \multicolumn{5}{l}{$\mathbf{8}(S=\frac{1}{2},\frac{3}{2},\frac{5}{2},\frac{7}{2})$}
\end{tabular}$,
$\begin{tabular}{|c|c|c|c|c|}
  \hline
  \quad \quad & \quad \quad & \quad \quad & \quad \quad & \quad \quad  \\
  \hline
  \quad \quad & \quad \quad  \\
  \cline{1-2}
  \quad \quad & \quad \quad \\
  \cline{1-2}
  \multicolumn{5}{c}{$\mathbf{10}(S=\frac{3}{2})$}
\end{tabular}$,
$\begin{tabular}{|c|c|c|c|}
  \hline
  \quad \quad & \quad \quad & \quad \quad & \quad \quad  \\
  \hline
  \quad \quad & \quad \quad & \quad \quad & \quad \quad \\
  \hline
  \quad \quad \\
  \cline{1-1}
  \multicolumn{4}{c}{$\mathbf{\bar{10}}(S=\frac{3}{2})$}
\end{tabular}$,
$\begin{tabular}{|c|c|c|c|c|}
  \hline
  \quad \quad & \quad \quad & \quad \quad & \quad \quad & \quad \quad  \\
  \hline
  \quad \quad & \quad \quad & \quad \quad \\
  \cline{1-3}
  \quad \quad \\
  \cline{1-1}
  \multicolumn{5}{c}{$\mathbf{27}(S=\frac{1}{2},\frac{3}{2},\frac{5}{2})$}
\end{tabular}$,
$\begin{tabular}{|c|c|c|c|c|c|}
  \hline
  \quad \quad & \quad \quad & \quad \quad & \quad \quad & \quad \quad & \quad \quad \\
  \hline
  \quad \quad & \quad \quad \\
  \cline{1-2}
  \quad \quad \\
  \cline{1-1}
  \multicolumn{6}{c}{$\mathbf{35}(S=\frac{1}{2})$}
\end{tabular}$,
$\begin{tabular}{|c|c|c|c|c|}
  \hline
  \quad \quad & \quad \quad & \quad \quad & \quad \quad & \quad \quad  \\
  \hline
  \quad \quad & \quad \quad & \quad \quad & \quad \quad \\
  \cline{1-4}
  \multicolumn{5}{c}{$\mathbf{\bar{35}}(S=\frac{1}{2})$}
\end{tabular}$,
$\begin{tabular}{|c|c|c|c|c|c|}
  \hline
  \quad \quad & \quad \quad & \quad \quad & \quad \quad & \quad \quad & \quad \quad \\
  \hline
  \quad \quad & \quad \quad & \quad \quad \\
  \cline{1-3}
  \multicolumn{6}{c}{$\mathbf{64}(S=\frac{3}{2})$ } \\
\end{tabular} $
\end{small} \\
The brackets below each flavor state show the possible spin states.

{\it SU(3) symmetric case:}

In Table~\ref{color-spin}, we show the $K$  values of the tribaryon for all allowed SU(3) flavor-spin states together with  the lowest threshold s-wave three baryon mode in the last row.   It should be noted that the $K$ values for the nucleon and Delta   are $-8$ and 8, respectively.  Therefore, using $\Delta K=16$  and $m_\Delta-m_p\sim 290$ MeV, we extract the constant factor in Eq.~(\ref{cs}) as $C_{CS}/m_u^2=18.125$ MeV.  As can be seen in the table, the difference of the color-spin interaction between the tribaryon and three octet baryon states are repulsive in all possible flavor and spin channel.  When the total spin is 9/2, the lowest threshold comes from three $\Delta$'s and the difference in $K$ vanishes.  This is similar to the two baryon case where the only non repulsive two baryon configuration is also in the maximum spin 3 lowest isospin 0 channel\cite{Park:2015nha}, which shows a possible dibaryon configuration\cite{Clement:2016vnl}.  However, the $\Delta$'s will decay
into a pion and a nucleon making the effective $K$ factor equal to the nucleon, in the massless pion limit, when the $\Delta$ appears in the lowest threshold baryon state as given in Eq.~(\ref{binding}).  This will effectively lower the $V$ value in Table~\ref{color-spin} by -16 for every $\Delta$ appearing as the lowest threshold baryon and contribute to the repulsion between the ground state baryons.  Such mechanism will take place whenever there is at least one $\Delta$ state in the lowest baryon threshold, which will be case for all configurations with $S \geq 5/2$.

For a realistic estimate, it is crucial to consider the SU(3) symmetry breaking effect\cite{SilvestreBrac:1992yg}.  The naive attraction or repulsion obtained by using the SU(3) symmetric formula in Eq.~(\ref{color-spin-formula}), can in some cases be quite misleading.  For example, the color-spin attraction in the H dibaryon against the two $\Lambda$ channel almost disappears when the SU(3) breaking effects are taken into account\cite{Park:2016cmg}, which also naturally explains the lattice results which show bound H dibaryons in the flavor SU(3) symmetric limit with massive pion\cite{Beane:2010hg,Inoue:2010es}.

Therefore, we investigate the effect of SU(3) breaking by calculating the color-spin matrix element together with their corresponding masses as given in Eq.~(\ref{cs}).  We then analyze Eq.~(\ref{binding}) as a function of the strange quark mass using a variable $\delta=1-\frac{m_u}{m_s}$.

\begin{table}
\begin{tabular}{|c|c|c|c|c|c|}
  \hline
  \multirow{2}{*}{Flavor} & \multicolumn{5}{|c|}{$-\sum_{i<j} \lambda_i \lambda_j \sigma_i \cdot \sigma_j$} \\
  \cline{2-6}
   & $S=\frac{1}{2}$ & $S=\frac{3}{2}$ & $S=\frac{5}{2}$ & $S=\frac{7}{2}$ & $S=\frac{9}{2}$ \\
  \hline
  $\mathbf{1}$ & & $-4$ & $\frac{8}{3}$ & & 24 \\
  \hline
  $\mathbf{8}$ &  4 & 8 & $\frac{44}{3}$ & 24 & \\
  \hline
  $\mathbf{10}$ & & 20 & & & \\
  \hline
  $\mathbf{\bar{10}}$ & & 20 & & & \\
  \hline
  $\mathbf{27}$ & 24 & 28 & $\frac{104}{3}$ & & \\
  \hline
  $\mathbf{35}$ & 40 & & & & \\
  \hline
  $\mathbf{\bar{35}}$ & 40 & & & & \\
  \hline
  $\mathbf{64}$ & & 56 & & & \\
  \hline
  $V$ & -24 & -24 & -8 & 8 & 24 \\
  \hline
\end{tabular}
\caption{The expectation value of color-spin interaction of the tribaryon for each flavor in SU(3) flavor symmetry. Empty boxes represent states that are not allowed by Pauli principle.  $V$ is for the lowest threshold three baryon states.}
\label{color-spin}
\end{table}

{\it Strangeness $=-1$:}
As discussed before, the case with one strange quark is of great interest in relation to hyperon properties in neutron star matter.
When there is only one strange quark, the flavor can not be
$\mathbf{1}$, $\mathbf{8}$ or $\mathbf{10}$.  Then, as can be seen from
Table~\ref{color-spin}, the least repulsive tribaryon configuration is the flavor  $\mathbf{\bar{10}}$ state.  Therefore, the eight non strange quark of the tribaryon configuration should be in the  $[44]_F$, which has isospin zero.  This flavor state of eight light quarks can also come from
$\mathbf{\bar{35}}$.
 For these states, the
the color-spin state of eight quarks should be $[2222]_{CS}$.
 Since eight quarks can have $S=4,3,2,1,0$, which correspond $[8]_S$,$[71]_S$,$[62]_S$,$[53]_S$ and $[44]_S$, we can find out which color-spin state is allowed using Clebsch-Gordan series of symmetric group $S_8$.
\begin{align}
  [332]_C \otimes [8]_S \Rightarrow &[332]_{CS} \nonumber\\
  [332]_C \otimes [71]_S \Rightarrow &[431]_{CS}, [422]_{CS}, [332]_{CS}, [3311]_{CS}, \nonumber\\
    & [3221]_{CS} \nonumber\\
  [332]_C \otimes [62]_S \Rightarrow &[53]_{CS}, [521]_{CS}, [431]_{CS}, [422]_{CS},  \nonumber\\
    & [4211]_{CS}, [332]_{CS}, [3311]_{CS}, [3221]_{CS},  \nonumber\\
    & [32111]_{CS}, [22211]_{CS} \nonumber\\
  [332]_C \otimes [53]_S \Rightarrow &[62]_{CS}, [53]_{CS}, [521]_{CS}, [5111]_{CS}, [44]_{CS}, \nonumber\\
    & [431]_{CS}, [422]_{CS}, [4211]_{CS}, [332]_{CS}, \nonumber\\
    & [3311]_{CS}, [3221]_{CS}, [2222]_{CS}, [41111]_{CS}, \nonumber\\
    & [32111]_{CS}, [22211]_{CS}, [221111]_{CS} \nonumber\\
  [332]_C \otimes [44]_S \Rightarrow &[611]_{CS}, [53]_{CS}, [521]_{CS}, [431]_{CS}, \nonumber\\
    & [4211]_{CS}, [332]_{CS}, [3221]_{CS}, [32111]_{CS},\nonumber\\
    & [22211]_{CS}, [311111]_{CS}
\end{align}

Therefore, the spin state of eight light quarks should be $[53]_S$, which is $S=1$. Since the spin state of eight quarks is determined, we can calculate the color-spin interaction value in $\mathrm{SU(3)}_F$ broken case easily using  Eq.~(\ref{color-spin-formula}). When there is only one strange quark with $I=0$, the expectation values of the color-spin interaction for $S=3/2$ and $S=1/2$, which  correspond to flavor $\mathbf{\bar{10}}$ and $\mathbf{\bar{35}}$ states respectively, are given as\\
\begin{itemize}
  \item{$S=\frac{3}{2}$}
  : $H_{SS}=20+\frac{20}{3}\delta$
  \item{$S=\frac{1}{2}$}
  : $H_{SS}=40-\frac{40}{3}\delta$,
\end{itemize}
where
\begin{eqnarray}
-\sum_{i<j} \frac{1}{m_i m_j} \lambda^c_i \lambda^c_j
{\sigma}_i\cdot{\sigma}_j   \equiv \frac{1}{m_u^2}H_{SS}.
\end{eqnarray}

As we can see in the Table~\ref{color-spin}, the color-spin interaction of the lowest decay mode is $-24$ for both S=3/2 and S=1/2 states, even when there is one strange quark. Therefore, we can conclude that the tribaryon with one strange quark is highly repulsive regardless of the strange quark mass value parametrized by $0<\delta<1$.

{\it Strangeness $=-3$:}
To investigate configurations with possible attraction, we will look at states with three strange quarks and isospin zero, which also include the most attractive tribaryon configuration as can be seen from Table I.
For these quantum numbers, there are four possible flavor states  $\mathbf{1}$, $\mathbf{8}$, $\mathbf{27}$ and $\mathbf{64}$.\\
\\
\begin{small}
$\begin{tabular}{|c|c|c|c}
  \cline{1-3}
  \quad \quad & \quad \quad & \quad \quad  \\
  \cline{1-3}
  \quad \quad & \quad \quad & \quad \quad \\
  \cline{1-3}
  $s$ & $s$ & $s$ \\
  \cline{1-3}
  \multicolumn{4}{c}{$\mathbf{1}(S=\frac{3}{2},\frac{5}{2},\frac{9}{2})$}
\end{tabular}$,
$\begin{tabular}{|c|c|c|c|c}
  \cline{1-4}
  \quad \quad & \quad \quad & \quad \quad & \hspace{0.08cm}$s$\hspace{0.08cm} \\
  \cline{1-4}
  \quad \quad & \quad \quad & \quad \quad  \\
  \cline{1-3}
  $s$ & $s$ \\
  \cline{1-2}
  \multicolumn{5}{l}{$\mathbf{8}(S=\frac{1}{2},\frac{3}{2},\frac{5}{2},\frac{7}{2})$}
\end{tabular}$,
$\begin{tabular}{|c|c|c|c|c|c}
  \cline{1-5}
  \quad \quad & \quad \quad & \quad \quad & \hspace{0.08cm}$s$\hspace{0.08cm} & \hspace{0.08cm}$s$\hspace{0.08cm}  \\
  \cline{1-5}
  \quad \quad & \quad \quad & \quad \quad \\
  \cline{1-3}
  \hspace{0.08cm}$s$\hspace{0.08cm} \\
  \cline{1-1}
  \multicolumn{6}{c}{$\mathbf{27}(S=\frac{1}{2},\frac{3}{2},\frac{5}{2})$}
\end{tabular}$,
$\begin{tabular}{|c|c|c|c|c|c|}
  \hline
  \quad \quad & \quad \quad & \quad \quad & \hspace{0.08cm}$s$\hspace{0.08cm} & \hspace{0.08cm}$s$\hspace{0.08cm} & \hspace{0.08cm}$s$\hspace{0.08cm} \\
  \hline
  \quad \quad & \quad \quad & \quad \quad \\
  \cline{1-3}
  \multicolumn{6}{c}{$\mathbf{64}(S=\frac{3}{2})$ } \\
\end{tabular}$
\end{small}\\

Here, we calculate the wave function of the tribaryon in the flavor SU(3) breaking case and fix the position of each quark as $q(1)q(2)q(3)q(4)q(5)q(6)s(7)s(8)s(9)$. The flavor-color-spin wave function should satisfy the antisymmetric property \{123456\}\{789\}. Then, we calculate the expectation value of the color-spin interaction for the possible tribaryon configurations. The results are as follows.
\begin{itemize}
  \item{$S=\frac{9}{2}$}
  : $H_{SS}=24-16\delta+8\delta^2$
  \item{$S=\frac{7}{2}$}
  : $H_{SS}=24-16\delta+8\delta^2$
\end{itemize}
\begin{widetext}
\begin{itemize}
\item{$S=\frac{5}{2}$} :
$
  H_{SS}=\left(
  \begin{array}{ccc}
   \frac{104}{3}-\frac{144}{5}\delta+\frac{44}{5}\delta^2 & -\frac{4\sqrt{14}}{5}\delta+\frac{2\sqrt{14}}{15}\delta^2 & \frac{4\sqrt{2}}{3\sqrt{5}}\delta^2 \\
   -\frac{4\sqrt{14}}{5}\delta+\frac{2\sqrt{14}}{15}\delta^2 & \frac{44}{3}-\frac{664}{45}\delta+\frac{374}{45}\delta^2 & \frac{8\sqrt{35}}{9}\delta+\frac{4\sqrt{7}}{9\sqrt{5}}\delta^2 \\
   \frac{4\sqrt{2}}{3\sqrt{5}}\delta^2 & \frac{8\sqrt{35}}{9}\delta+\frac{4\sqrt{7}}{9\sqrt{5}}\delta^2 & \frac{8}{3}-\frac{16}{9}\delta+\frac{80}{9}\delta^2 \\
  \end{array}
  \right) \nonumber
$
\item{$S=\frac{3}{2}$} :
$
  H_{SS}=\left(
  \begin{array}{cccc}
    56-\frac{368}{7}\delta+\frac{200}{21}\delta^2 & -\frac{32\sqrt{5}}{21}\delta+\frac{16}{21\sqrt{5}}\delta^2 & -\frac{8}{\sqrt{105}}\delta^2 & 0 \\
    -\frac{32\sqrt{5}}{21}\delta+\frac{16}{21\sqrt{5}}\delta^2 & 28-\frac{2824}{105}\delta+\frac{1843}{210}\delta^2 & -\frac{14\sqrt{7}}{5\sqrt{3}}\delta+\frac{\sqrt{3}}{5\sqrt{7}}\delta^2 & \frac{7}{2\sqrt{15}}\delta^2 \\
    -\frac{8}{\sqrt{105}}\delta^2 & -\frac{14\sqrt{7}}{5\sqrt{3}}\delta+\frac{\sqrt{3}}{5\sqrt{7}}\delta^2 & 8-\frac{88}{15}\delta+\frac{128}{15}\delta^2 & \frac{2\sqrt{35}}{3}\delta+\frac{\sqrt{7}}{3\sqrt{5}}\delta^2 \\
    0 & \frac{7}{2\sqrt{15}}\delta^2 & \frac{2\sqrt{35}}{3}\delta+\frac{\sqrt{7}}{3\sqrt{5}}\delta^2 & -4+\frac{8}{3}\delta+\frac{55}{6}\delta^2 \\
  \end{array}
  \right) \nonumber
$
\item{$S=\frac{1}{2}$} :
$
  H_{SS}=\left(
  \begin{array}{cc}
  24-\frac{80}{3}\delta+\frac{28}{3}\delta^2 & \frac{4\sqrt{2}}{3}\delta+\frac{2\sqrt{2}}{3}\delta^2 \\
  \frac{4\sqrt{2}}{3}\delta+\frac{2\sqrt{2}}{3}\delta^2 & 4-8\delta+\frac{26}{3}\delta^2 \\
  \end{array}
  \right) \nonumber
$
\end{itemize}
\end{widetext}

For each possible spin states, we subtract the lowest three baryon threshold as given in Eq.~(\ref{binding})  and plot them  in Fig.~\ref{color-spin}.
 For SU(3) flavor mixed configurations, the diagonalized ground states are plotted.
\begin{figure}[htbp]
  \begin{center}
         \includegraphics[scale=0.8]{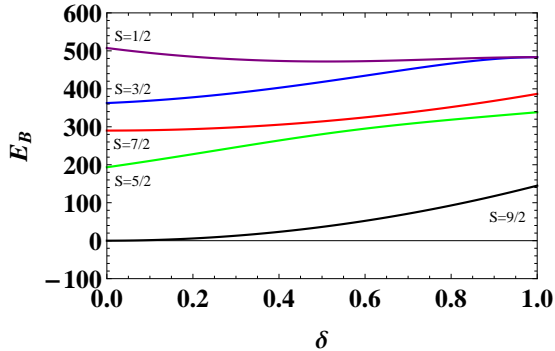}
\caption[]{$E_B$ of $q^6 s^3$ with $I=0$(unit: MeV).}
\label{color-spin}
  \end{center}
\end{figure}
Again, as can be seen in the figure, the tribaryon configuration are highly repulsive against three lowest baryon channel except for the spin 9/2 channel, which does not contribute to the three octet baryon channel as they appear only up to spin 3/2.

One can also consider intrinsic quark three body interaction.  However,  one can show that for color singlet configurations of $N$ quarks, the two possible terms add to
\begin{eqnarray}
\sum_{i \neq j\neq k} f^{abc} \lambda^a_i \lambda^b_j \lambda^c_k & = & 0 \nonumber \\
\sum_{i \neq j\neq k} d^{abc} \lambda^a_i \lambda^b_j \lambda^c_k & = &-8 NC_1(q) \bigg(2C_1(q)-\frac{13}{3}\bigg) ,
\end{eqnarray}
where $f,d$ are the antisymmetric and symmetric structure constants for SU(3), respectively, and $C_1(q)$ the first Casimir of the quarks. Hence, as in the case of color-color interaction, the only non-vanishing term is also proportional to $N$ and does not contribute to repulsion.

We have shown that the static compact three baryon configurations are all highly repulsive with respect to three baryon channel.
While the model is based on constituent quark model, we have analyzed all possible two and three body quark interaction and at the distance scale that is valid for the constituent quark model.
This is the first microscopic and clear proof  based on a constituent quark model that the three body nuclear force should be repulsive at short distance in all channels in the SU(3) limit.  We have further analyzed and shown that the repulsion remains valid even when the SU(3) breaking  effect is taken into account in the most attractive YNN and YYY channels, which are consistent with constraints from nuclear many body observables and the recently established neutron star mass limit.


\section*{Acknowledgments}
This work was supported by the Korea National Research
Foundation under the grant number 2016R1D1A1B03930089, 2018R1D1A1B07043234, and 2017R1D1A1B03028419(NRF).

\end{document}